\documentclass[runningheads]{llncs}
\usepackage[T1]{fontenc}
\usepackage{graphicx}
\usepackage{algorithmic}
\usepackage{amssymb}
\usepackage{algorithm}
\usepackage{amsmath}
\usepackage{amsfonts}
\usepackage{color}
\usepackage{siunitx}
\usepackage{booktabs}
\usepackage{threeparttable}
\usepackage{multirow}
\usepackage{marvosym}
\usepackage{hyperref}
\usepackage{color}

\begin{document}
\title{InsMix: Towards Realistic Generative Data Augmentation for Nuclei Instance Segmentation}
\titlerunning{InsMix: Towards Realistic Generative Data Augmentation.}
\authorrunning{Lin \textit{et al.}, InsMix}
\author{Yi Lin\textsuperscript{(\Letter)}, Zeyu Wang, Kwang-Ting Cheng, Hao Chen}
\institute{Department of Computer Science and Engineering\\ The Hong Kong University of Science and Technology, Kowloon, Hong Kong\\
\email{linyi.pk@gmail.com}}
\maketitle              
\begin{abstract}
Nuclei Segmentation from histology images is a fundamental task in digital pathology analysis. 
However, deep-learning-based nuclei segmentation methods often suffer from limited annotations. 
This paper proposes a realistic data augmentation method for nuclei segmentation, named InsMix, that follows a Copy-Paste-Smooth principle and performs morphology-constrained generative instance augmentation.
Specifically, we propose morphology constraints that enable the augmented images to acquire luxuriant information about nuclei while maintaining their morphology characteristics (e.g., geometry and location).
To fully exploit the pixel redundancy of the background and improve the model's robustness, we further propose a background perturbation method, which randomly shuffles the background patches without disordering the original nuclei distribution.
To achieve contextual consistency between original and template instances, a smooth-GAN is designed with a foreground similarity encoder (FSE) and a triplet loss.
We validated the proposed method on two datasets, i.e., Kumar and CPS datasets. 
Experimental results demonstrate the effectiveness of each component and the superior performance achieved by our method to the state-of-the-art methods.
\footnote{The source code is available at \url{https://github.com/hust-linyi/insmix}.}
\end{abstract}
\section{Introduction}
Nuclei segmentation is a crucial step for the analysis of computational pathology images. 
Cancer diagnosis and treatment are directly influenced by the distribution and morphology (e.g., size, shape, and location) of the nuclei~\cite{elmore2015diagnostic}.
Recent advances in deep learning~\cite{litjens2017survey,zhou2020deep,lin2022label} have led to remarkable success in nuclei instance segmentation tasks. 
For example, Micro-Net~\cite{liao2016automatic} utilized multi-resolutions and a weighted loss function to achieve robustness against the large inter-/intra-variability of the nuclei size.
To separate the touching/overlapping nuclei, some studies~\cite{chen2020boundary,cui2019deep,xie2020instance,zhou2019cia,chen2017dcan} incorporate the boundary information of the nuclei into the segmentation task. 
For example, TAFE~\cite{chen2020boundary} aggregated multi-scale information into two separate branches; one for the nuclei boundary and the other for the nuclei content.
Some studies~\cite{graham2019hover,naylor2018segmentation} proposed to regress the distance map of nuclei to avoid predicting areas with indistinguishable boundaries.
Some other methods tried to integrate nuclei detection and segmentation into one single network~\cite{zhou2019irnet} where the segmentation results are used to refine the nuclei detection.

\begin{figure}[!t]
    \centering
    \includegraphics[width=\textwidth]{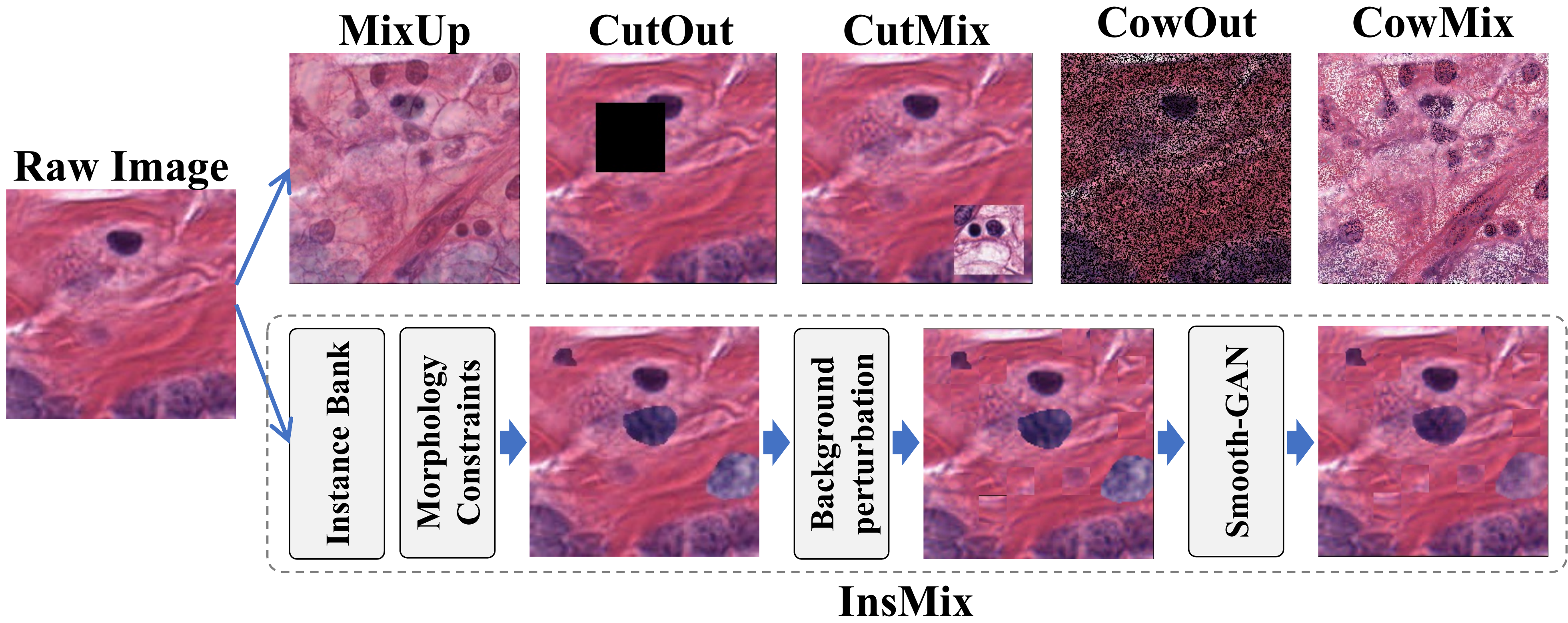}
    \caption{Illustration of Mix-based data augmentation methods, and our InsMix method.}
    \label{fig_visual_mix}
\end{figure}

In spite of these advances in nuclei instance segmentation, one fundamental challenge of the current deep-learning-based methods is a lack of sufficient amount of annotated data for training.
Accurate pixel-wise annotation of the nuclei,  which requires clinical expertise, is a labor-intensive and time-consuming procedure.
Data augmentation is the most straightforward way to overcome this limitation. 
In addition to the conventional data augmentation methods (e.g., flipping and rotation), several Mix-based methods have been proposed.
For instance, as shown in Fig.~\ref{fig_visual_mix}, the MixUp method~\cite{zhang2018mixup} is a data augmentation method that combines random samples of the input images.
CutOut~\cite{devries2017improved} augments an image by randomly masking a rectangular region to zero.
CutMix~\cite{yun2019cutmix} incorporates MixUp and CutOut that randomly crops a patch from one image and places it onto another one.
CowOut and CowMix were proposed by French et al.~\cite{french2019semi} that extends CutOut and CutMix by introducing a random cropping mask.

Copy-Paste~\cite{dvornik2018modeling,dwibedi2017cut} is another way to combine multiple images' information, which can be viewed as a form of object-level CutMix, where foreground pixels from one image are copied and pasted onto another.
The effectiveness of this type of method for the instance segmentation task has been successfully validated. 
In particular, Ghiasi et al.~\cite{ghiasi2021simple} performed a systematic study of the Copy-Paste method, and achieved state-of-the-art results on COCO instance segmentation~\cite{lin2014microsoft} and LVIS benchmarks~\cite{gupta2019lvis}.
Nevertheless, there are still some limitations of this method, especially its application to nuclei instance segmentation.
First, the Copy-Paste method simply copies instances from one image to another, which may result in some loss of clinical prior information, such as nuclei distribution and location.
Second, the Copy-Paste method could result in obvious irregular appearance, due to the significant color variation between the original and template instances caused by staining.

To address these challenges, this paper proposes a novel data augmentation method, named InsMix, which performs Copy-Paste-Smooth, achieving more realistic data augmentation.
The main differences of InsMix from the previous Copy-Paste methods~\cite{dwibedi2017cut,fang2019instaboost,ghiasi2021simple,xu2021continuous} are at least in the following three aspects: 1) Instead of directly performing Copy-Paste, we propose morphology constraints (SSD, i.e., scale, shape, and distance) to maintain nuclei's morphology characteristics (i.e., location, clustering, etc);
2) In addition to foreground augmentation, we propose a background perturbation method to fully exploit effective use of the background information for data augmentation and in turn strengthen the robustness of the segmentation model;
3) To generate realistic augmented images, we introduce smooth-GAN based on a triplet loss, where we design a foreground similarity encoder (FSE) to encode the original nuclei contextual information into the template nuclear instances. 
Extensive experiments on Kumar~\cite{kumar2017dataset} and CPS~\cite{zhou2019irnet} datasets show the proposed data augmentation methods substantially improve the nuclei instance segmentation performance compared with state-of-the-art techniques.

\section{Method}
As shown in Fig.~\ref{fig_framework}, our InsMix aims at realistic instance augmentation in a Copy-Paste-Smooth manner. Specifically, the foreground instances are augmented under the morphology constraints, named SSD constraints. Then a background perturbation method is proposed to fully exploit the background information. Last, a smooth-GAN is proposed to eliminate the artifacts of template instances. In the following, we elaborate on each component in detail.

\subsection{Foreground Augmentation with Morphology Constraints}
\begin{figure}[!t]
    \centering
    \includegraphics[width=\textwidth]{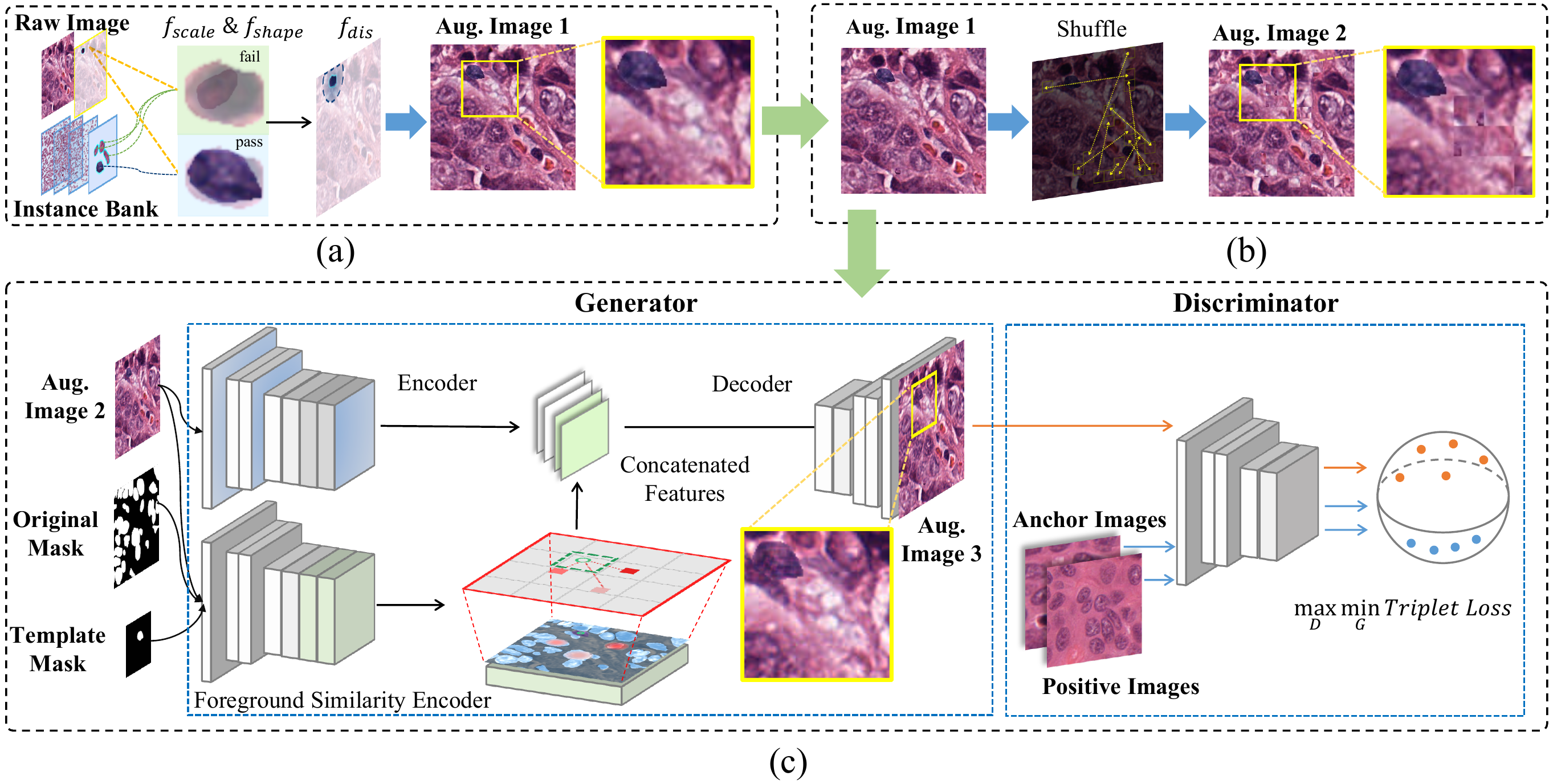}
    \caption{Illustration of (a) morphology constraints, (b) background perturbation, and (c) smooth-GAN.}
    \label{fig_framework}
\end{figure}
As mentioned above, the Copy-Paste~\cite{dwibedi2017cut} method may lack the rationality of the pasted instances~\cite{fang2019instaboost} by randomly copying the foreground from one image to another.
To this end, we propose SSD constraints considering the morphology characteristics of the original instances, as shown in Fig.~\ref{fig_framework}(a).
Specifically, for more feasible augmentation, we first construct an instance bank by collecting all the instances from the training set as the templates. 
In this way, we can flexibly control the number of pasted instances, regarding the number of original instances.
Then, for each instance $I_o$ of the original image $\mathcal{I}_O$, and each template instance $I_t \in \mathcal{I}_T$, we apply the SSD constraints $f_{\mathrm{SSD}}(I_o, I_t)$ as follows:
\begin{equation}
f_{\mathrm{SSD}}(I_o, I_t):\ f_{scale}(I_o, I_t) \le \epsilon;\  f_{shape}(I_o, I_t) \le \rho;\  \delta \le f_{dis}(I_o, I_t) \le \gamma,
\end{equation}
where $f_{scale}$, $f_{shape}$, and $f_{dis}$ represent the functions for evaluating the scale difference, shape consistency, and centroid distance, respectively. The parameters $\epsilon$, $\rho$, $\delta$ and $\gamma$ are determined by cross-validation. Given the binary mask $M_{o}$ and $M_{t}$ of $I_o$ and $I_t$, the function $f_{scale}$ and $f_{shape}$ can be formulated as:
\begin{equation}
f_{scale}(I_o, I_t) = \frac{\max(|M_o|, |M_t|)}{\min(|M_o|, |M_t|)}, f_{shape}(I_o, I_t) = \frac{|M_o - M_t|}{\max(|M_o|, |M_t|)}, 
\end{equation}
larger $f_{scale}$ and $f_{shape}$ indicate greater scale and larger shape inconsistency, respectively. 
To ensure that the position of template nuclei follows the original instance distribution, we further restrict the centroid distances of the two masks by $\delta$ and $\gamma$. 
By changing $\delta$, we can easily obtain touching/overlapping nuclei instances; and by changing $\gamma$, the template instances can locate into the surrounding region of the target instance, avoiding clinically meaningless results.

\subsection{Background Perturbation for Robustness Improvement}
In practice, a common preprocessing step in nuclei segmentation is to split large pathology image into smaller patches. 
This may cause ambiguity regarding the sharp edges of the patches. 
Based on this observation, we introduce the background perturbation method to randomly shuffle the background patches. 
Specifically, as shown in Fig.~\ref{fig_framework}(b) we first split the background region into $20\times20$ patches, and then randomly shuffle the patches with the ratio of $\alpha$ (which is empirically set to 0.2). In this way, the nuclei distribution would not be disordered, and the segmentation model would be robust to the distraction of the sharp edges, such as the irregular shape and incomplete texture of nuclei.

\subsection{Smooth-GAN for Realistic Instance Augmentation}
Due to the various stainings in the training set, there exists an obvious color shift between the template and original instances  (as shown in Fig.~\ref{fig_visual_mix}).
Thus, we introduce smooth-GAN to generate realistic and smoothed results.
As shown in Fig.~\ref{fig_framework}(c), our smooth-GAN adopts the image-to-image translation method~\cite{pix2pix2017}, which consists of a generator and a discriminator. 
The generator is to translate the augmented images with a smoother appearance and boundary by borrowing the contextual information from the original instances to the template instances.
The discriminator is to distinguish the unrealistic augmented images. 

\noindent\textbf{Training the Discriminator.}
We use the PatchGAN~\cite{pix2pix2017} with spectral normalization for the discriminator following~\cite{zeng2021cr}. 
Unlike conventional GAN-based methods, there's no ground-truth for the smoothed augmented images. 
We therefore employ the triplet loss to train the discriminator.
Specifically, we randomly select two raw images from the training set as the anchor and the positive image (i.e., $x_a$ and $x_p$), and the negative image is the smoothed result from the generator.
The discriminator is trained by the following triplet loss with $l_1$ distance:
\begin{equation}
\begin{aligned}
\mathcal{L}_D = \mathbb{E}_{x\sim P_{\mathrm{ori}}, u\sim P_{\mathrm{aug}}}(0, &\left|D(x_a) - D(x_p)\right|  \\
& - \left|D(x_a) - D\left(G(u)\circ M + u\circ (1-M)\right)\right| + m),
\end{aligned}
\end{equation}
where $D(\cdot)$ and $G(\cdot)$ denote the discriminator and generator, respectively. $M$ denotes the binary mask corresponding to the template instance's region where $M_{xy}=1$ indicates that pixel at $[x,y]$ belongs to the template instance. $\circ$ represents element-wise multiplication. The smoothed result $S(u) = G(u)\circ M + u\circ (1-M)$ is composed by putting the generated template instance $G(u)$ in the original image while keeping the other region of $u$. The discriminator is trained to narrow down the perceptional distance between the anchor and positive samples and enlarge the distance between the anchor and negative samples with the margin $m$ (which is empirically set to 1.0 in our experiments).

\noindent\textbf{Training the Generator.}
The generative network adopts a typical auto-encoder network~\cite{pix2pix2017}. 
We employ gated convolution~\cite{zeng2021cr} and dilated convolution in the network for a large receptive ﬁeld~\cite{yu2018generative}. 
The adversarial loss is defined as:
\begin{equation}
\mathcal{L}_{adv} = \mathbb{E}_{x\sim P_{\mathrm{ori}},u\sim P_{\mathrm{aug}}}[\left|D(x_a) - D\left(S(u)\right)\right| - \left|D(x_a)-D(x_p)\right|],
\end{equation}
In addition, for maintaining the original image's information, we add the reconstruction loss (i.e., $l_1$ loss) to the adversarial loss as the final generative loss:
\begin{equation}
\mathcal{L}_G = \mathcal{L}_{adv} + \lambda |u- G(u)|,
\end{equation}
where $\lambda$ is the weight parameter, which is empirically set to 10 in our experiment.

\noindent\textbf{Foreground Similarity Encoder (FSE).}
The generator, however, would collapse that it simply learns to apply background redundancy appearance to template instances' regions.
To overcome this problem, we introduce an auxiliary encoder, named foreground similarity encoder (FSE), which borrows the appearance information (e.g, staining) from the original instances to the template instances.
We first calculate the feature similarity between the original and template instances, and then integrate the original instance feature into the template feature space. 
As shown in Fig.~\ref{fig_framework}(b), we first extract the $3\times 3$ patches in the original instances' region. 
Then for each patch in the template instances' region, we calculate the cosine similarity for the original instance patches in a convolutional way as: 
\begin{equation}
\mathcal{S}_{i, j} = \frac{f(p_i)^\intercal s(p_j)}{|f(p_i)| \cdot |f(p_j)|}, i\in I_t, j\in I_o,
\end{equation}  
where $f(p_i)$ and $f(p_j)$ is the feature of the patches in the template and original instance's region, respectively. 
Then, we replace the template instance feature with the original instance feature, weighted by the normalized similarity as follows:
\begin{equation}
\bar{f}(p_i) = \sum_{j\in \mathcal{I}_A} \mathrm{softmax}(\mathcal{S}_{i, j}) \cdot f(p_j).
\end{equation}
The obtained similarity encoding $\bar{f}(u)$ is then concatenated with the raw encoder feature as input for the decoder to generate the final augmented image. In this way, the augmented images would achieve global appearance consistency.

\section{Experiments}
\noindent\textbf{Datasets and Implementation Details.}
The Kumar dataset~\cite{kumar2017dataset} contains 30 H\&E stained pathology images with a resolution of $1000\times 1000$ pixels.
We follow the same dataset splitting criterion as~\cite{chen2020boundary} that 16 images for training and 14 images for testing.
In the test set, 8 images are from the same organs as the training set (denoted as seen organ), and 6 images are from 3 organs that are not in the training set (unseen organ).

For quantitative evaluation, we use Dice coefficient (Dice) and aggregated Jaccard index (AJI) as the evaluation metrics, which are the most commonly used evaluation metrics in nuclei instance segmentation at the pixel and object level, respectively. Note that this work focus on instance-level data augmentation, thus AJI would be  more suitable for this work.
During training, we crop $256\times 256$ patches from the raw images. 
For a fair comparison, instead of the proposed InsMix augmentation, we also apply traditional data augmentation methods for all the compared methods, e.g., randomly scale, shift, rotation, flip, color jittering, and blurring.
We train the model in a total of 300 epochs, using the Adam optimizer with weight decay $10^{-4}$. 
The initial learning rate is set to $10^{-3}$ with the cosine annealing schedule, resetting in every 50 epochs. 

\begin{table}[htbp]
  \centering
  \caption{Comparison with the state-of-the-art methods on Kumar.}
  \label{tab_sota}
  \setlength{\tabcolsep}{2mm}{
    \begin{tabular}{l|ccc|ccc}
    \toprule
    \multirow{2}{*}{Methods} & \multicolumn{3}{c|}{Dice (\%)} & \multicolumn{3}{c}{AJI (\%)} \\
    \cline{2-7} 
            & Seen  & Unseen    & All   & Seen  & Unseen    & All \\
    \hline
    CNN3~\cite{kumar2019multi}    & 82.26    & 83.22    & 82.67 & 51.54    & 49.89    & 50.83    \\
    DIST~\cite{naylor2018segmentation}    & -         & -         & - & 55.91    & 56.01    & 55.95    \\
    NB-Net~\cite{cui2019deep}    & 79.88    & 80.24    & 80.03 & 59.25    & 53.68    & 56.86    \\
    Mask R-CNN~\cite{he2017mask}  & 81.07    & 82.91    & 81.86 & 59.78    & 55.31    & 57.86    \\
    HoVer-Net~\cite{graham2019hover} (*Res50)   & 80.60     & 80.41    & 80.52   & 59.35    & 56.27    & 58.03    \\
    TAFE~\cite{chen2020boundary} (*Dense121)       & 80.81    & 83.72    & 82.06   & 61.51    & 61.54    & 61.52    \\
    \hline
    HoVer-Net $+$ InsMix  & 80.33 & 81.93 & 81.02 & 59.40 & 57.67 & 58.66 \\ 
    TAFE $+$ InsMix   & \textbf{81.18}    & \textbf{84.40}    & \textbf{82.56}   & \textbf{61.98}    & \textbf{65.07}    & \textbf{63.31}    \\
    \bottomrule
    \end{tabular}}
\end{table}

\noindent\textbf{Comparison with SOTA.}
We compare the proposed method with the state-of-the-art methods~\cite{chen2020boundary,graham2019hover,kumar2019multi,naylor2018segmentation,zhou2019cia} on the Kumar datasets.
For the methods without public codes, we report the results from the original publications for a fair comparison.
As depicted in Table~\ref{tab_sota}, promising results in nuclei instance segmentation results are observed using the proposed data augmentation methods. Our InsMix significantly boosts the AJI by 2.23\% and 0.63\%, compared with TAFE~\cite{chen2020boundary} and HoVer-Net~\cite{graham2019hover}, respectively. 
And Our InsMix sightly improves the Dice score by 0.5\%.
We argue that our InsMix exploits the nuclei knowledge at the instance level, which hence improves the AJI by a large margin.

\begin{figure}[!t]
    \centering
    \includegraphics[width=\textwidth]{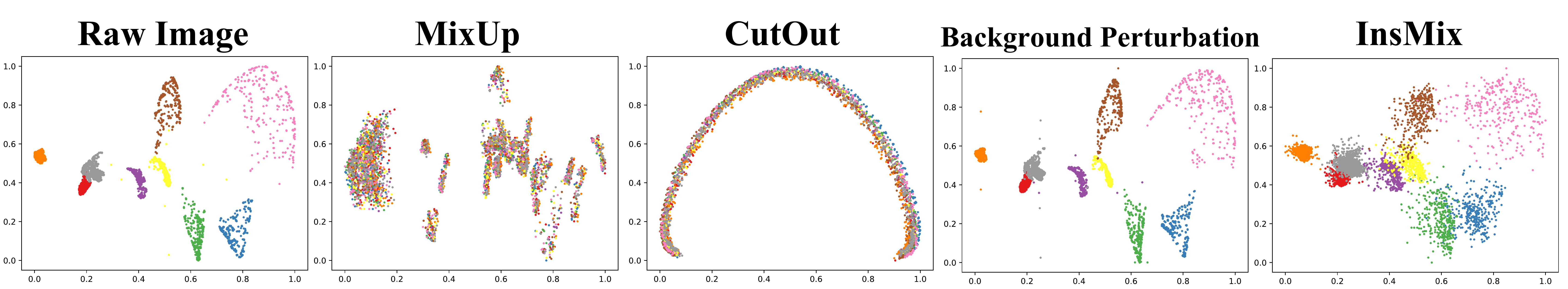}
    \caption{T-SNE embedding of image patches with different data augmentations. Different colors indicate cropped patches from different images. 
    }
    \label{visual_tsne}
\end{figure}

\begin{table}[!t]
  \centering
  \caption{Comparison with other Mix-based methods on Kumar.}
  \label{tab_related}
  \setlength{\tabcolsep}{1mm}{
    \begin{tabular}{c|cccccc}
    \toprule
                & MixUp~\cite{zhang2018mixup} & CutOut~\cite{devries2017improved} & CutMix~\cite{yun2019cutmix} & CowOut~\cite{french2019semi} & CowMix~\cite{french2019semi} & InsMix \\
    \hline
    Dice (\%)   & 81.21 & 82.01 & 82.33 & 82.27 & 81.80 & \textbf{82.56} \\   
    AJI (\%)    & 61.68 & 62.29 & 61.61 & 62.87 & 61.19 & \textbf{63.31}\\   
    \bottomrule
    \end{tabular}}
\end{table}

For comprehensive evaluation, we further compared our InsMix methods with other Mix-based data augmentation methods, including MixUp~\cite{zhang2018mixup}, CutOut~\cite{devries2017improved}, CutMix~\cite{yun2019cutmix}, CowOut~\cite{french2019semi}, and CowMix~\cite{french2019semi}. 
The results are shown in Table~\ref{tab_related}. 
It can be seen that our InsMix outperforms other methods by 0.44\%-2.12\% in AJI. 
We believe that the improvement comes from the full exploration of both foreground and background information, without introducing distraction of the data distribution. 
A more intuitionistic explanation can be found in Fig.~\ref{visual_tsne}, where we conduct t-SNE for the augmented images. It can be seen that our InsMix could fill up the low density in data distribution, without introducing undesirable bias (e.g., CutOut) or distractions (e.g., MixUp).

\begin{table}[!t]
  \centering
  \caption{Ablation study on the Kumar dataset, where SSD, s-GAN, BgP denote SSD constraints, smooth-GAN, and background perturbation, respectively.}
  \label{tab_ablation}
  \setlength{\tabcolsep}{1pt}{
    \begin{tabular}{c|c|c|c|c|c}
    \toprule
                & Baseline    &  $+$SSD   & $+$BgP   &  $+$SSD$+$s-GAN  &  $+$SSD$+$BgP$+$s-GAN\\ 
    \hline
    AJI (\%)    & 61.52       & 62.08  & 62.24   & 62.37   & \textbf{63.31} \\
    \bottomrule
    \end{tabular}}
\end{table}

\noindent\textbf{Ablation Study.}
To validate each component in the InsMix method, we perform an ablation study on Kumar, taking TAFE~\cite{chen2020boundary} as the baseline.
The results are presented in Table~\ref{tab_ablation}.
The baseline method achieves an AJI of 61.52\%, and the SSD constraints boost the AJI to 62.08\%. 
We believe this is because the SSD constraints could take clinical prior into account to generate more meaningful results regarding the scale, shape, and distribution of the nuclei, which is different from the simple Copy-Paste method.
Smooth-GAN could further improve the AJI to 62.37\%. 
The visualization results of the smooth-GAN are shown in Fig.~\ref{fig_visual_mix}.
Lastly, with background perturbation, the AJI increases to 63.31\%.

\begin{figure}[!t]
    \centering
    \includegraphics[width=\textwidth]{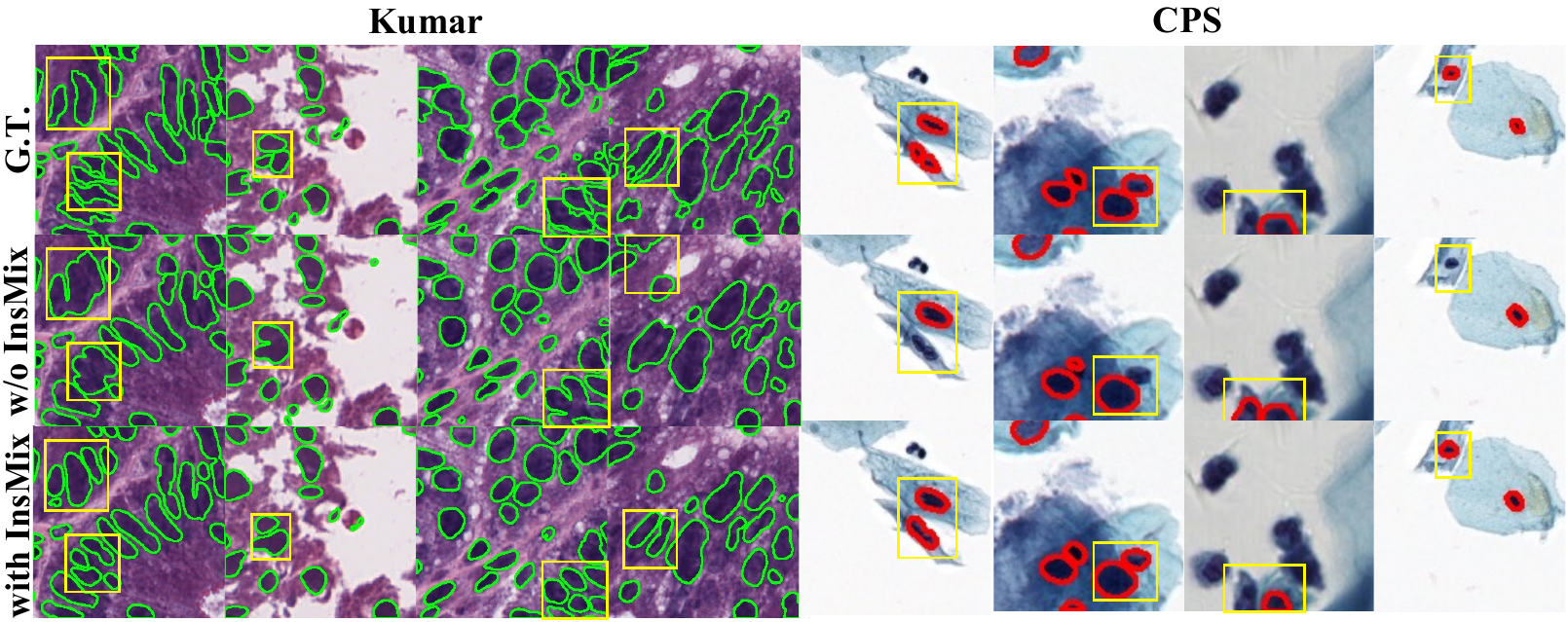}
    \caption{Instance segmentation results on Kumar and CPS. Rectangles highlight differences among different methods.
    }
    \label{visual_cps}
\end{figure}

\begin{table}[!t]
  \centering
  \caption{Comparison with other Mix-based methods on CPS. A 5-fold cross-validation is conducted.}
  \label{tab_cps}
   \resizebox{\textwidth}{!}{
    \begin{tabular}{c|ccccccc}
    \toprule
                & Baseline & MixUp~\cite{zhang2018mixup} & CutOut~\cite{devries2017improved} & CutMix~\cite{yun2019cutmix} & CowOut~\cite{french2019semi} & CowMix~\cite{french2019semi} & InsMix \\
    \hline
    Dice (\%)   & 69.29$\pm$4.41  & 69.83$\pm$3.87  & 69.79$\pm$4.16 & 68.81$\pm$3.03  & 61.91$\pm$10.46 & 69.40$\pm$4.30 & \textbf{70.96}$\pm$4.59 \\   
    AJI (\%)    & 49.45$\pm$4.44 & 49.83$\pm$3.82 & 50.23$\pm$4.00 & 48.48$\pm$2.43 & 39.77$\pm$9.98 & 48.06$\pm$4.79 & \textbf{51.54}$\pm$3.93\\   
    \bottomrule
    \end{tabular}}
\end{table}

\noindent\textbf{Validation on Other Dataset.}
To evaluate the generality of the proposed method, we experiment on the cervical Pas smear (CPS) image dataset~\cite{zhou2019irnet}. The dataset contains 82 Pap smear images with the size of $1000\times 1000$ pixels. 
We adopt NB-Net~\cite{cui2019deep} with ResUNet-34 as the baseline.
We perform 5-fold cross-validation to evaluate the effectiveness of the proposed method, compared with baseline and other Mix-based methods.
In Table~\ref{tab_cps}, on average 5 folds, the InsMix improves the performance of the baseline method by 1.67\% in Dice and 2.09\% in AJI, respectively. Our InsMix also outperforms other Mix-based methods.

\section{Conclusion}
In this paper, we propose a novel data augmentation method, named InsMix, for nuclei instance segmentation. 
Our InsMix applies morphology constraints (SSD, i.e., scale, shape, and distance) to maintain the clinical nuclei priors.
Besides foreground augmentation, we also propose background perturbation to exploit the pixel redundancy of the background.
Further, a smooth-GAN is proposed to uniform the contextual information between the original and template nuclei. 
Experimental results demonstrated the effectiveness of each component and the superior performance of our model to the state-of-the-art methods.

\bibliographystyle{splncs04}
\bibliography{ref}

\end{document}


\title{InsMix: Towards Realistic Generative Data Augmentation for Nuclei Instance Segmentation}
\titlerunning{InsMix: Towards Realistic Generative Data Augmentation.}
\authorrunning{Lin \textit{et al.}, InsMix}
\author{Yi Lin\textsuperscript{(\Letter)}, Zeyu Wang, Kwang-Ting Cheng, Hao Chen}
\institute{Department of Computer Science and Engineering\\ The Hong Kong University of Science and Technology, Kowloon, Hong Kong\\
\email{linyi.pk@gmail.com}}
\maketitle              

\begin{table}[thbp]
    \centering
    \caption{Dice (\%) and AJI (\%) on CPS. A 5-fold cross-validation is conducted.}
    \label{tab:my_label}
    \resizebox{\textwidth}{!}{
    \begin{tabular}{c|ccccc>{\columncolor{mycyan}}c|ccccc>{\columncolor{mycyan}}c}
    \toprule
\multirow{2}{*}{Methods}  &\multicolumn{6}{c|}{Dice} &\multicolumn{6}{c}{AJI} \\ 
\cline{2-13}
        &fold1  &fold2  &fold3  &fold4  &fold5  &Avg.   &fold1  &fold2  &fold3  &fold4  &fold5  &Avg. \\
\hline
baseline	&70.87	&70.77	&74.73	&66.96	&63.13	&69.29 		&51.49	&50.1	&54.86	&47.87	&42.94	&49.45 \\
mixup	&70.62	&71.57	&74.96	&66.56	&65.46	&69.83 	&51.02	&51.72	&54.5	&46.82	&45.1	&49.83 \\
cutout	&72.21	&70.89	&74.83	&65.31  &65.72  &69.79  &52.12	&49.99	&55.97	&46.56	&46.5	&50.23 \\
cutmix	&70.68	&70.84	&71.52	&65.43	&65.58	&68.81  &50.98	&50.92	&48.08	&45.48	&46.95	&48.48 \\
cowout	&64.79	&64.34	&71.68	&44.01	&64.72	&61.91 	&42.06	&41.77	&49.86	&23.00	&42.17	&39.77 \\
cowmix	&67.13	&72.76	&75.21	&65.7	&66.22	&69.40 	&42.76	&51.66	&54.39	&45.51	&45.99	&48.06 \\
insmix	&72.52	&73.43	&76.43	&66.85	&65.58	&\textbf{70.96} 	&51.89	&52.96	&57.11	&48.78	&46.95	&\textbf{51.54} \\
\bottomrule
    \end{tabular}}
\end{table}

\begin{figure}[htbp]
    \centering
    \caption{The visualization results of smooth-GAN.}
    \includegraphics[width=\textwidth]{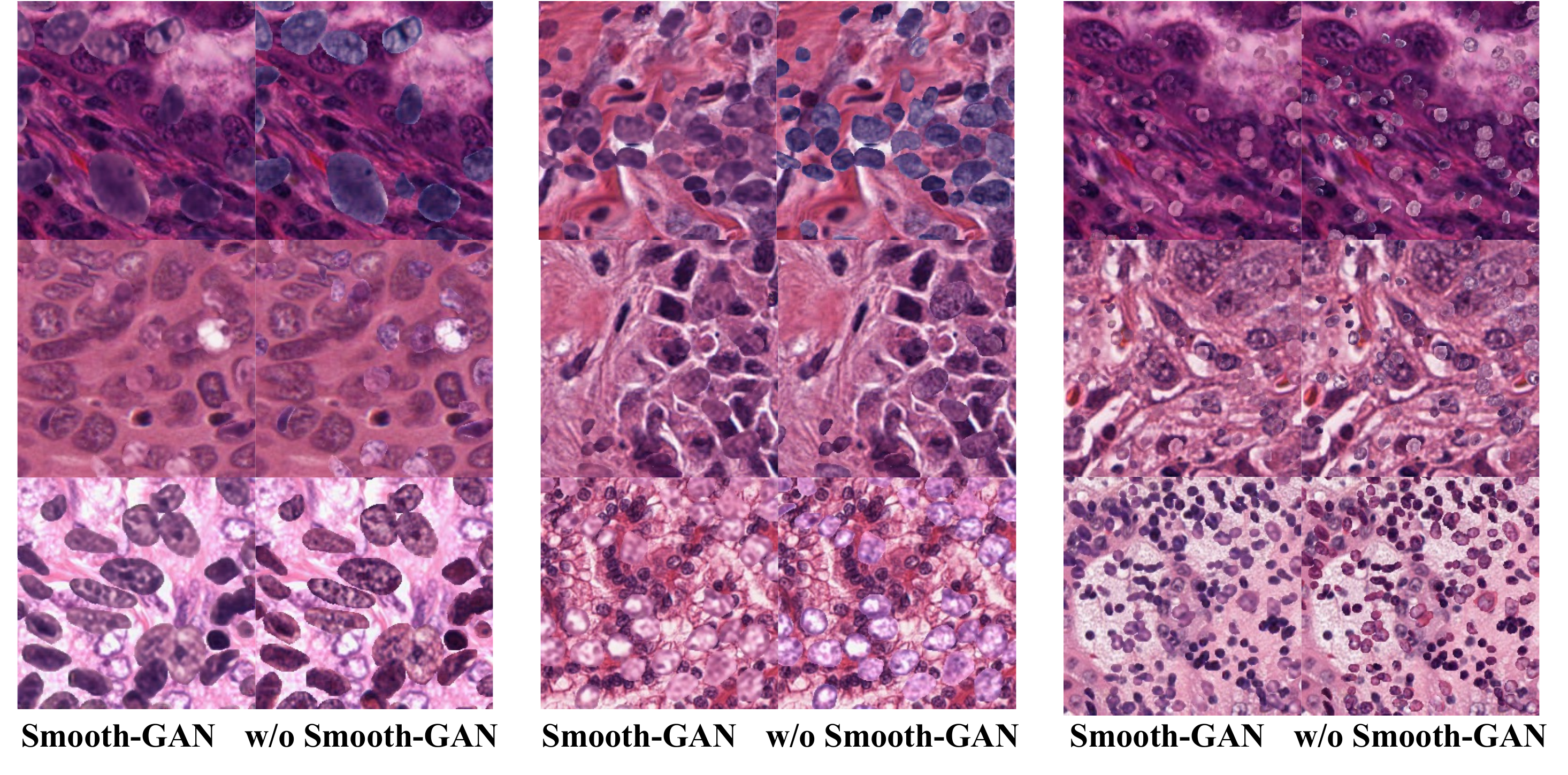}
\end{figure}

\begin{figure}[htbp]
    \centering
    \caption{The visualization results of Kumar.}
    \includegraphics[width=\textwidth]{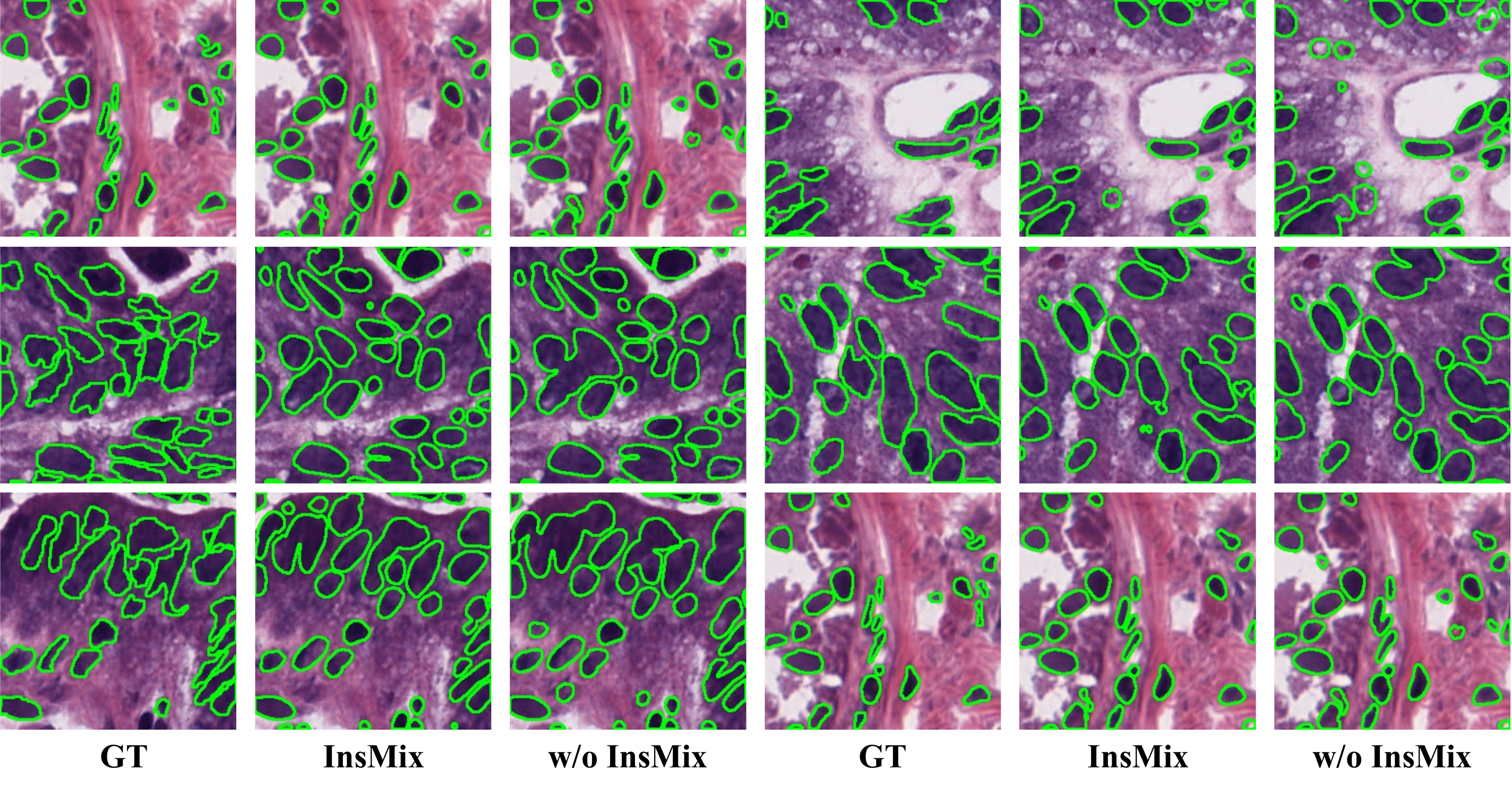}
\end{figure}

\begin{figure}[htbp]
    \centering
    \caption{The visualization results of CPS.}
    \includegraphics[width=\textwidth]{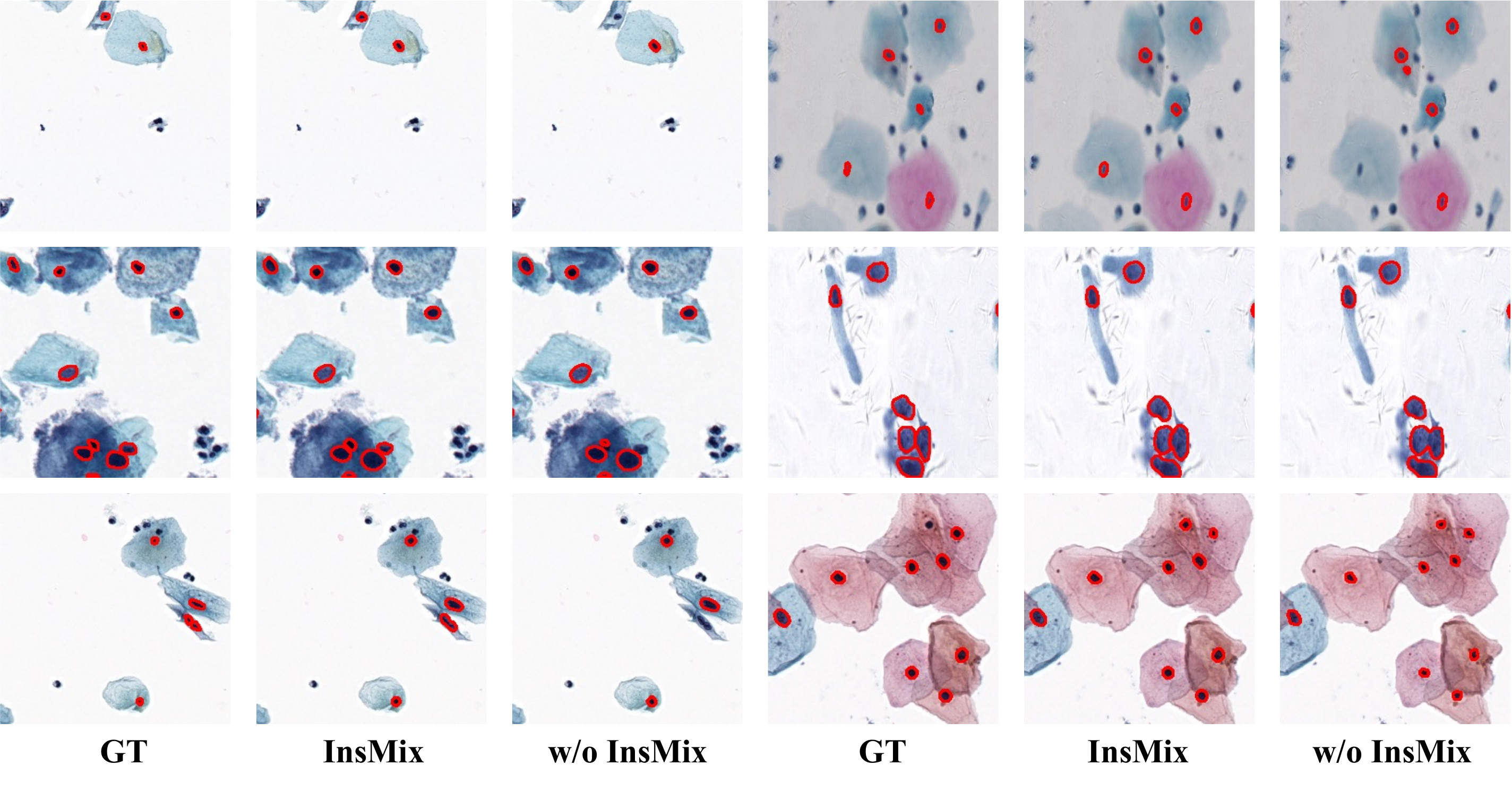}
\end{figure}